\documentclass[aps,prl,showpacs,twocolumn]{revtex4}
\newcommand{\bk}{{\bf k}}
\newcommand{\br}{{\bf r}}
\newcommand{\lan}{\langle}
\newcommand{\ran}{\rangle}
\newcommand{\bj}{{\bf J}}
\newcommand{\trho}{\tilde\rho}
\newcommand{\dr}{\Delta\br}
\newcommand{\bv}{{\bf v}}
\newcommand{\pnm}{P_m^{(0)}}
\newcommand{\pnj}{P_j^{(0)}}
\newcommand{\hti}{{\hat H}}
\newcommand{\gan}{\gamma_n}
\newcommand{\st}{{\tilde\sigma}}
\newcommand{\stil}{{\tilde s}}

\newcommand{\sgt}{{\tilde G}_0}
\newcommand{\ovd}{\overline{\overline{d}}}

\begin{document}
\title{Hopping models and ac universality}
\author{Jeppe C. Dyre}\email{dyre@ruc.dk}
\author{Thomas B. Schr{\o}der}
\affiliation{Department of Mathematics and Physics (IMFUFA), 
Roskilde University, Postbox 260, DK-4000 Roskilde, Denmark}
%\date{\today}
\begin{abstract}
Some general relations for hopping models are established. We proceed  to
discuss the universality of the ac conductivity which arises in the extreme
disorder limit of the random barrier model. It is shown that the relevant
dimension entering into the diffusion cluster approximation (DCA) is the
harmonic (fracton) dimension of the diffusion cluster. The temperature
scaling of the dimensionless frequency entering into the DCA is discussed.
Finally, some open questions about ac universality are mentioned.
\end{abstract}
\maketitle

\section{Introduction}
Hopping models have been studied for many years in different
contexts \cite{shk84,bot85,hau87,bou90}. Our main concern here is
the modeling of ac conduction in disordered solids, a field where
hopping models are quite successful. There is a wealth of ac data
for various disordered solids like ionic conductive glasses,
amorphous semiconductors, polymers, etc. These solids behave 
similarly; from bulk ac data alone it is not possible to distinguish ionic
and electronic conduction. The ``universal'' ac features \cite{dyr00} are:
At low frequencies the conductivity is constant, at high frequency it
follows an approximate power law with an exponent which is less than one but
slowly approaches one as frequency increases. As temperature is lowered the
dc conductivity goes to zero and the frequency exponent goes to one (at any
fixed frequency much below phonon frequencies). The frequency marking onset
of ac conduction is proportional to the dc conductivity, while both are
strongly temperature dependent (usually Arrhenius). 

In this paper we first establish hopping models as part of a more
general framework. We then proceed to discuss the extreme disorder
limit of hopping models. In this limit the ac conductivity in
properly scaled units becomes independent of the jump frequency
probability distribution. Finally, new results are established for
the diffusion cluster approximation, a recently proposed
analytical approximation for the universal ac conductivity.

\section{diffusion and hopping models}

First, we briefly consider diffusion in general and define the wave-number
and frequency-dependent diffusion constant.
Then we specialize to hopping of non-interacting particles on a
cubic lattice and derive a number of relations. Many of the points
made below can be found in the literature, but to our knowledge they
have never been concisely written down aimed at hopping models.

If $\rho$ is the particle density and $\bf J$ the particle current
density, the most general diffusion-type constitutive equation obeying
linearity, causality and space-time homogeneity is:
\begin{equation}\label{1}
\bj (\br,t)\ =\ -\int d\br' \int_0^\infty dt'\ D(\br',t') {\bf
\nabla}\rho (\br-\br',t-t')\,.
\end{equation}
It is convenient to introduce the so-called Laplace frequency $s$
given by 
\begin{equation}\label{1a}
s\ =\ i\omega\,,
\end{equation}
where $\omega$ is the ordinary frequency. The wave-number and
frequency-dependent diffusion constant, $D(\bk,s)$, is by definition the
Fourier-Laplace transform of $D(\br',t')$:
\begin{eqnarray}\label{2}
D(\bk,s)\ &\equiv &\ 
\int d\br'\ e^{-i\bk\cdot\br'}\int_0^\infty dt'\ e^{-st'}
D(\br',t')\ \nonumber\\ &\equiv & 
\int d\br'\ e^{-i\bk\cdot\br'} D(\br',s)\,.
\end{eqnarray}
It is possible to express $D(\bk,s)$ in terms of equilibrium density
fluctuations. To show this we first Laplace transform the equation of
continuity, $\dot{\rho}=-\nabla\cdot\bf
J$, and calculate the right hand side by means of Eq. (\ref{1}).
If $\trho$ is the Laplace transform of $\rho$ this leads to
\begin{equation}\label{3}
s\trho(\br,s)-\rho(\br,0)\ =\
\int d\br' D(\br',s) \nabla^2\trho(\br-\br',s)\,.
\end{equation}
We proceed by Fourier transforming this equation. If $\rho_\bk$ is
the Fourier transform of the density, $\rho_\bk=\int
d\br\rho(\br)\exp(-i\bk\cdot\br)$, one finds
\begin{equation}\label{4}
s\trho_\bk(s)-\rho_\bk(0)\ =\ - D(\bk,s) k^2 \trho_\bk(s)\,.
\end{equation}
Equation\ (\ref{4}) implies
$\trho_\bk(s)=\rho_\bk(0)/\left[s+D(\bk,s)k^2\right]$. Multiplying
this relation by $\rho_{-\bk}(0)$ we find after thermally
averaging
\begin{equation}\label{5}
\frac{\lan\trho_\bk(s)\rho_{-\bk}(0)\ran}
{\lan\rho_\bk(0)\rho_{-\bk}(0)\ran}\ =\ 
\frac{1}{s+D(\bk,s)k^2}\,.
\end{equation}

We are mainly interested in the $\bk\rightarrow 0$ limit. The
frequency-dependent diffusion constant $D(s)$ is defined by 
\begin{equation}\label{6}
D(s)\ =\ \lim_{\bk\rightarrow 0} D(\bk,s)\,.
\end{equation}
We proceed to derive an equation relating $D(s)$ to the mean-square
displacement as function of time, considering from now on only the
case of {\it non-interacting} particles. If $d$ is space dimension and
$\dr(t)$ is particle displacement in time $t$, the relevant expression is 
(with an implicit convergence factor 
$\lim_{\epsilon\rightarrow 0}\exp(-\epsilon t)$, $\epsilon>0$):
\begin{equation}\label{7}
D(s)\ =\ 
\frac{s^2}{2d}\int_0^\infty\lan\dr^2(t)\ran\ e^{-st}dt\,.
\end{equation}
The proof proceeds as follows. If the particle positions are
denoted by $\br^{(j)}$ we have 
$\rho_\bk(t)=\sum_j \exp\left[-i\bk\cdot\br^{(j)}(t)\right]$.
Non-interacting particles move uncorrelated and thus
$\lan\rho_\bk(t)\rho_{-\bk}(0)\ran=
\sum_j\lan\exp\left[-i\bk\cdot\dr^{(j)}(t)\right]\ran$. Consequently,
it is enough to consider the motion of just one particle and
\begin{equation}\label{8}
\frac{\lan\rho_\bk(t)\rho_{-\bk}(0)\ran}
{\lan\rho_\bk(0)\rho_{-\bk}(0)\ran}\ =\ 
\lan e^{-i\bk\cdot\dr(t)}\ran\,.
\end{equation}
Being only interested in the limit of small $\bk$, the exponential
on the right hand side is expanded:
$\exp\left(-i\bk\cdot\dr\right)
=1-i\bk\cdot\dr-\frac{1}{2}(\bk\cdot\dr)^2+...$. For the thermal
average we have by symmetry $\lan\dr\ran=0$ and thus
$\lan\exp\left(-i\bk\cdot\dr\right)\ran
=1-\frac{1}{2}\lan(\bk\cdot\dr)^2\ran+...$. Spherical symmetry
implies that, if $\Delta r_i$ is the $i$'th coordinate of $\dr$ 
etc, $\lan\Delta r_i\Delta r_j\ran=\lan\dr^2\ran\delta_{ij}/d$.
Thus
$\lan(\bk\cdot\dr)^2\ran=\lan k_i\Delta r_i k_j \Delta r_j\ran =
k^2\lan\dr^2\ran/d$. Substituting this into Eq.\ (\ref{8}), Laplace
transforming, and utilizing Eq.\ (\ref{5}) leads to
\begin{equation}\label{9}
\frac{1}{s+D(\bk,s)k^2}\ =\ 
\frac{1}{s}-\frac{k^2}{2d}\int_0^\infty\lan\dr^2(t)\ran
e^{-st}dt+...\,.
\end{equation}
For small $\bk$ the left hand side is expanded as follows: 
$s^{-1}\left(1-s^{-1}D(\bk,s)k^2+...\right)$. Comparing this to
the right hand side and letting $\bk\rightarrow 0$ proves Eq.\
(\ref{7}).

The case of ``ordinary'' diffusion is characterized by
$\lan\dr^2(t)\ran=2dDt$ where $D$ is the ordinary diffusion
constant. Substituting this into Eq.\ (\ref{7}) gives $D(s)=D$ at
all frequencies. Thus the definition of $D(s)$ is a consistent
generalization of the ordinary diffusion constant. Finally, we
note that Eq.\ (\ref{7}) may be rewritten by performing two
partial integrations:
\begin{equation}\label{10}
D(s)\ =\ 
\frac{1}{d}\int_0^\infty\lan\bv(0)\cdot\bv(t)\ran e^{-st}dt\,.
\end{equation}

Henceforth we specialize to the case of non-interacting particles
hopping on a finite cubic lattice in $d$ dimensions. Hopping
models have been discussed for several years in the literature.
Some reviews of hopping are listed in Refs.
\cite{shk84,bot85,hau87,bou90}, some recent interesting papers
dealing with ac hopping are given in Refs.
\cite{bar99,por00,rol01}. In order to be able to speak about well-defined
displacements we shall {\it not} impose periodic boundary conditions. This
implies $D(0)=0$. To calculate $D(0)$ in the physically realistic
``bulk'' limit one should calculate $D(s)$ at a non-zero value of
$s$, let lattice size go to infinity and only thereafter let $s$
go to zero. Alternatively, one imposes periodic boundary
conditions and doesn't consider positions at all but velocities,
and calculates $D(s)$ by means of Eq.\ (\ref{10}) \cite{sch99}.

As mentioned already, it is only necessary to consider the motion of one
particle. A hopping model is defined as follows. Adopting the bra-ket
notation of quantum mechanics the lattice sites are denoted by $|j\ran$. Any
state of the system $|P\ran$ is given by an expression of the form
$|P\ran=\sum_m P_m|m\ran$, where $P_m$ is the probability to find
the particle at site $m$. The state space is a real Hilbert space
when equipped with ordinary inner product but, of course, only
states with positive probabilities summing to $1$ are physical.
Time development is described by a master equation which is
conveniently written in terms of a ``Hamiltonian'' as follows: 
\begin{equation}\label{11}
\frac{d}{dt}\ |P\ran\ =\ 
H\ |P\ran\,.
\end{equation}
Here, $H$ is the well-known master equation time-development
matrix constructed from the transition probabilities. $H$ has the
following properties \cite{van81}: All diagonal elements of $H$
are non-positive (giving the decay rate from a state), all
off-diagonal elements are non-negative (giving the transition rate
from one state to another). The condition $\sum_jH_{jm}=0$ ensures
probability conservation. Finally, if $\pnm$ is the canonical probability to
find a particle at site $m$ in thermal equilibrium, for any states $|j\ran$
and $|m\ran$ the principle of detailed balance demands that
$H_{jm}\pnm=H_{mj}\pnj$. 

Since the solution of Eq.\ (\ref{11}) is 
$|P(t)\ran=\exp(Ht)|P(0)\ran$ the mean-square displacement in
thermal equilibrium is given by
\begin{equation}\label{12}
\lan\dr^2(t)\ran\ =\ \sum_{jm}\left(\br_j-\br_m\right)^2
\lan j| e^{Ht} | m\ran\ \pnm\,.
\end{equation}
As always when dealing with a master equation it is convenient to
switch to the ``symmetric'' representation \cite{van81}: First we
define the operator $T$ by
$T|m\ran=\sqrt{\pnm}|m\ran\equiv|\psi_m\ran$. It is easy to show
that the principle of detailed balance implies that the operator
$\hti\equiv T^{-1}HT$ is Hermetian (real and symmetric).
Furthermore, $\lan j|\exp(Ht)|m\ran=\lan j|\exp\left(T\hti
T^{-1}t\right)|m\ran=\lan j|T\exp(\hti
t)T^{-1}|m\ran=\lan\psi_j|\exp(\hti t)|\psi_m\ran/\pnm$.
Substituting this into Eq.\ (\ref{12}) leads to
\begin{equation}\label{13}
\lan\dr^2(t)\ran\ =\ 
\sum_{jm}\left(\br_j-\br_m\right)^2
\lan\psi_j|e^{\hti t}|\psi_m\ran\,.
\end{equation}

From this important conclusions are arrived at. First, note that
if any state may be reached from any other - ``ergodicity'' - $\hti$ has a
unique ``ground state'' with eigenvalue $0$
\cite{van81}. This state, $|0\ran$, is the symmetric representation of the
state of thermal equilibrium. It is
straightforward to show that this normalized eigenstate is given
by $|0\ran=\sum_m|\psi_m\ran$. All other eigenvalues of $\hti$,
denoted by $-\gan$, are strictly negative (i.e., $\gan >0$). If the
corresponding eigenstates of $\hti$ are denoted $|n\ran$, the usual trick of
``sandwiching in'' the orthonormal set of eigenstates leads to
$\lan\psi_j|\exp(\hti t) |
\psi_m\ran=\pnj\pnm+\sum_n\lan\psi_j|n\ran\lan n|\psi_m\ran
\exp(-\gan t)$. Substituting this into Eq.\ (\ref{13}) leads to
\begin{equation}\label{14}
\lan\dr^2(t)\ran\ =\ 
\sum_{jm}\left(\br_j-\br_m\right)^2\pnj\pnm
-\sum_n \mu_n e^{-\gan t}\,,
\end{equation}
where $\mu_n\equiv -\sum_{jm}\left(\br_j-\br_m\right)^2
\lan\psi_j|n\ran\lan n|\psi_m\ran$. Since $\lan\dr^2(t)\ran=0$ for
$t=0$ Eq.\ (\ref{14}) may be rewritten as
\begin{equation}\label{15}
\lan\dr^2(t)\ran\ =\ \sum_n \mu_n \left(1-e^{-\gan t}\right)\,.
\end{equation}
We proceed to prove that $\mu_n\geq 0$. Suppose that
$\lambda_j$ are real numbers summing to zero. Then
$\sum_{jm}\left(\br_j-\br_m\right)^2\lambda_j\lambda_m\leq 0$,
because the sum may be rewritten as
$\left(\sum_j\lambda_j\br_j^2\right)\left(\sum_m\lambda_m\right)
+\left(\sum_m\lambda_m\br_m^2\right)\left(\sum_j\lambda_j\right)
-2\left(\sum_j\lambda_j\br_j\right)^2$ and the first two terms
vanish. It follows that $\mu_n\geq 0$ because for any $|n\ran$
one has $\sum_j\lan\psi_j|n\ran=\lan 0|n\ran=0$.

We are now able to prove a result for the mathematical structure of
the mean-square displacement in hopping models: By repeated
differentiations Eq. (\ref{15}) implies
\begin{eqnarray}\label{16}
\frac{d}{dt}\ \lan\dr^2(t)\ran\ & \geq & 0\nonumber \\
\frac{d^2}{dt^2}\ \lan\dr^2(t)\ran\ & \leq & 0\\
 ... &{\rm etc.}& ...\nonumber
\end{eqnarray}
This is the first important consequence of Eq.\ (\ref{13}). A similar result
applies for $D(s)$. Substituting Eq.\ (\ref{15}) into Eq.\
(\ref{7}) leads to
\begin{eqnarray}\label{17}
D(s)\ & = & \ 
\frac{s^2}{2d}\sum_n\mu_n\left(\frac{1}{s}
-\frac{1}{s+\gan}\right)\nonumber\\
& = & \frac{1}{2d}\sum_n\mu_n\frac{s\gan}{s+\gan}\,.
\end{eqnarray}
Obviously, $D(0)=0$ and $D(s)\geq 0$ for $s\geq 0$. Equation (\ref{17}) is
further simplified
by writing in the numerator $s=s+\gan-\gan$, leading to
\begin{equation}\label{18}
D(s)\ =\ D(\infty)\ -\ \frac{1}{2d}
\sum_n\mu_n \frac{\gan^2}{s+\gan}\,,
\end{equation}
where $D(\infty)=\sum_n\mu_n\gan/2d$\,. It is straightforward to
show that $D(\infty)=\sum_{jm}(\br_j-\br_m)^2\Gamma_{jm}\pnm/2d$
where $\Gamma_{jm}$ is the transition rate from site $|m\ran$ to
site $|j\ran$ ($\Gamma_{jm}=-H_{jm}$ for $j\neq m$).
Differentiating Eq.\ (\ref{18}) with respect to s repeatedly for $s\geq 0$
leads to
\begin{eqnarray}\label{19}
\frac{d}{ds}\ D(s) & \geq & 0\nonumber \\
\frac{d^2}{ds^2}\ D(s) & \leq & 0\\
 ... &{\rm etc.}& ...\nonumber
\end{eqnarray}
Substituting $s=i\omega$ into Eq.\ (\ref{18}) immediately gives a result
first derived by Kimball and Adams \cite{kim78}: The real part of the
diffusion constant is always an increasing
function of $\omega$. 

Our final general result relates to the case $s > 0$, where the approximate
power law exponent obeys:
\begin{equation}\label{20}
0\ \leq\ \frac{d\ln D}{d\ln s}\ \leq \ 1\,.
\end{equation}
The first inequality simply expresses the fact that $D(s)$ is an
increasing function of s (Eq.\ (\ref{19})). Now, if $f_1(s)$ and
$f_2(s)$ are two positive functions each obeying $d\ln f/d\ln s\leq 1$
then this condition is also obeyed for $f_1+f_2$: We have
$sf_1'/f_1\leq 1$ or $f_1'\leq f_1/s$ and similarly for $f_2$. Adding
these two inequalities leads to $f_1'+f_2'\leq (f_1+f_2)/s$ or $d\ln
(f_1+f_2)/d\ln s\leq 1$. Next, we note that the function
$f(s)=Cs/(s+\gamma)$ obeys $0\leq d\ln f/d\ln s=\gamma/(s+\gamma)\leq 1$
whenever $C\geq 0$ and $\gamma\geq 0$. The second inequality Eq.\
(\ref{20}) now follows from Eq.\ (\ref{17}).

\section{Universality of the ac conductivity in the extreme
disorder limit}

If the particles are charged the fluctuation-dissipation theorem tells us
that the frequency-dependent electrical conductivity $\sigma(s)$ is
proportional to $D(s)$. In particular, in hopping models there are relations
for $\sigma(s)$ analogous to Eqs.\ (\ref{19}) and (\ref{20}), and the real
part of the conductivity, $\sigma'$, is an increasing function of $\omega$.

We now specialize to hopping models with only nearest-neighbor
jumps, and focus on the simplest model of this kind, the random barrier
model (also called the symmetric hopping model). For this model all sites on
the cubic lattice have equal
energy. Thus, detailed balance implies for the jump rates
$\Gamma_{jm}=\Gamma_{mj}$. Because we want to model a {\it
disordered} solid the jump rates are taken to vary randomly and
uncorrelated from link to link. Moreover, we shall assume that the
jump rates are given by a free energy barrier $E$:
$\Gamma=\Gamma_0\exp(-\beta E)$. Here $\beta$ is the inverse
temperature if classical barrier hopping is considered and inverse
wave function size if quantum mechanical tunneling is considered.
The model is completely defined in terms of the free energy barrier
probability distribution, $p(E)$. 

At large values of $\beta$ the jump rates vary many orders of
magnitude. The limit $\beta\rightarrow\infty$ is termed the {\it
extreme disorder limit}. In this limit, although all jump
frequencies go to zero and consequently $\sigma\rightarrow 0$, one
may ask how the conductivity {\it relative} to the dc level,
$\st\equiv\sigma/\sigma(0)$, behaves as a function of frequency
\cite{dyr94,sum85,van91}.
It turns out that $\st$ as function of a suitably scaled frequency in the
extreme disorder limit becomes independent of the barrier distribution. This
result, ``ac universality,'' applies for any smooth
$p(E)$ which is non-zero at the percolation threshold. More
accurately, ac universality means that for any fixed scaled
[Laplace] frequency $\stil$, as $\beta\rightarrow\infty$ the
conductivity $\st(\stil)$ converges to some value which is independent of
$p(E)$. In computer simulations we find that the larger $\stil$
is, the larger $\beta$ must be before there is convergence. Since
$\stil\sim 1$ defines the region below which the conductivity is
virtually frequency independent, this means that the convergence
to universality is fastest around the onset of ac conduction.

Although no mathematically rigorous proof of ac universality
exists, there is convincing evidence for ac universality from
several sources \cite{dyr00}. First of all, ac universality is
clearly seen in computer simulations, where several quite
different $p(E)$'s lead to the same $\st(\stil)$ in the
extreme disorder limit. Secondly, the effective medium
approximation (EMA) implies ac universality in the extreme
disorder limit with the following prediction \cite{dyr94}:
\begin{equation}\label{21}
\st\ \ln\st\ =\ \stil\,.
\end{equation}
This equation was first derived by Bryksin as the EMA solution of
the hopping model describing electrons tunneling between random
positions in space \cite{bry80}. Finally, it is possible to
physically understand the origin of universality; basically it stems from
the fact that {\it percolation} dominates
conduction in the extreme disorder limit \cite{dyr00}.

The EMA universality equation (\ref{21}) has the correct
qualitative features of the universal ac conductivity: a constant
low frequency conductivity and an ac conductivity which at high
frequencies follows an approximate power law; the exponent
is below one but converges logarithmically to one as
frequency diverges. Quantitatively, however, Eq.\ (\ref{21}) is
not accurate in three dimensions, and even less accurate in two
dimensions. In both cases the onset of ac conduction is less 
dramatic than predicted by Eq.\ (\ref{21}). One may speculate, however, that
Eq.\ (\ref{21}) becomes exact in and above 6 dimensions, because here
mean-field theory for percolation is exact (when regarded as a critical
phenomenon), and EMA is a sort of mean-field theory. We have preliminary
simulation data for 4 dimensions indicating that Eq.\ (\ref{21}) indeed
works better as dimension is increased.

Before proceeding to a discuss a more accurate analytical
approximation to the universal ac conductivity, let us briefly
sketch how Eq.\ (\ref{21}) is derived because this is relevant for
the following. The units used are ``rationalized units'' where
conductivity, diffusion constant and jump rate are all identical in
the ordered (homogeneous) case. The general EMA equation for finite disorder
is a rather complicated self-consistency equation for $\sigma(s)$
\cite{hau87}. This equation involves the diagonal element of the Green's
function for a ``homogeneous'' random walk with uniform jump frequency
$\Gamma$, $\sgt\equiv\lan\br|G_0|\br\ran$, where $G_0=1/(s-H_0)$ is the
resolvent operator for the uniform-case ``Hamiltonian'' $H_0$. $\sgt$ is a
function of $\Gamma(=\sigma)$, besides of course also a function of $s$. In
the extreme disorder limit the EMA self-consistency equation reduces
\cite{dyr94} to
\begin{equation}\label{22}
\ln\st\ \propto\ \beta\ s\sgt\,.
\end{equation}
It is straightforward to show that $s\sgt\propto s/\sigma$ for
small $s$ in two and more dimensions \cite{dyr94}. When this is
substituted into Eq.\ (\ref{22}) one arrives at Eq.\ (\ref{21})
after a suitable rescaling of $s$ to define the dimensionless
quantity $\stil$:

\begin{equation}\label{22a}
\stil\ =\ f(\beta)\frac{s}{\sigma(0)}\,,
\end{equation}
where $f(\beta)\propto\beta$.

As mentioned already Eq.\ (\ref{21}) does not give an accurate
representation of the universal ac conductivity in three
dimensions. A better fit to data is provided by what we term
the ``diffusion cluster approximation'' (DCA), which leads
\cite{sch00} to
\begin{equation}\label{23}
\ln\st\ =\ \left(\frac{\stil}{\st}\right)^{d_0/2}\,.
\end{equation}
Our simulations in 3-d are well fitted by $d_0=1.35$. An equation similar to
Eq.\ (\ref{23}) was derived by Zvyagin in 1980 \cite{zvy80} by
reference to cluster size statistics at percolation. Zvyagin's equation,
however, is real: He has $\omega$ where we have $i\omega$ and ${\rm
Re}(\st)$ where we have $\st$. 

In the original derivation of Eq.\ (\ref{23}) $d_0$ was identified with
the dimension of the ``diffusion cluster'' \cite{sch00}. This set is defined
as follows. We first recall that at extreme disorder conduction takes place
on the percolation cluster
\cite{amb71,shk71,kir73}. More precisely, links with jump rates
much smaller than the ``percolation jump rate'' can be removed
without affecting the overall (dc) conductivity of the lattice;
for any finite $\beta$ this leaves us with the ``fat'' percolation
cluster \cite{dyr00} (which becomes the true percolation cluster
as $\beta\rightarrow\infty$). 
However, more links may be removed. First of all, those on dead
ends of the percolation cluster may be removed, leaving the so-called
backbone. Moreover, if there are two different paths between any two points
below the correlation length, because of the extreme disorder one of them is
much more favorable than the other which may be removed. After this 
diluting of the percolation cluster the remainder is by definition the
diffusion cluster. Now, our basic assumption is that
\begin{itemize}
\item \textbf{In the extreme disorder limit not only dc but also 
ac conduction takes place mainly on the diffusion cluster.}
\end{itemize}
This assumption runs contrary to the traditional use of percolation
theory to calculate $\sigma(s)$. The assumption should be checked by
computer simulations, but we have not yet done so. Please note that, at any
finite $\beta$ the assumption applies only at not too high frequencies;
there are
always isolated islands of well-conducting regions which are
important at sufficiently high frequencies. Our conjecture, however, is
that the scaled frequency below which the conjecture applies
diverges as $\beta\rightarrow\infty$. 

What is the structure of the diffusion cluster? Let us first
recall the ``nodes-links-blobs'' model of the backbone. According to this
model \cite{nak94} the backbone comprises links (i.e., quasi
one-dimensional strings) and nodes at the intersection of links.
The ``blobs'' refer to the fact that for $0-1$ percolation there
are occasional ``strongly bonded'' regions along any link. However, as
argued above, blobs are unimportant in the extreme disorder limit
of a continuum distribution of jump rates because, of any two
different paths between two points one will strongly dominate. 

Our picture is now the following: At any finite $\beta$, the time
scale on which the particle moves more than the node-node distance
corresponds to frequencies where the conductivity is frequency
independent; here we agree with the standard use of percolation theory to ac
phenomena. On a smaller distance scale the diffusion cluster is
fractal. In our original derivation of Eq.\ (\ref{23}) we used 
the EMA equation Eq.\ (\ref{22}) for hopping on the
diffusion cluster and identified $d_0$ with the dimension of
the diffusion cluster, having in mind the fractal dimension
$\overline{d}$. More correctly, $d_0$ should be identified with
the so-called harmonic or fracton dimension $\ovd$, which for
homogeneous (uniform jump rate) random walks on the diffusion cluster gives
the probability $P(t)$ to be at the same place as at $t=0$: By definition
$P(t)\propto t^{-\ovd/2}$ and $\sgt$ is basically the Laplace
transform of P(t), which implies $s\sgt\propto(s/\sigma)^{\ovd/2}$. The
connection between $\ovd$ and $\overline{d}$ \cite{ale82,nak94} is
\begin{equation}\label{24}
\ovd \ =\ 
\frac{2\overline{d}}{2+\delta}\,,
\end{equation}
where the exponent $\delta$ is given by $\lan\dr^2(t)\ran\propto
t^{2/(2+\delta)}$ for a homogeneous random walk on the diffusion cluster.

Is it possible that $\overline{d}=\ovd$, corresponding to $\delta=0$? In our
opinion the answer is yes, because $\delta=0$
for any fractal without dead ends and without loops, e.g., a
selfsimilar curve like the Koch curve \cite{nak94,avm}. If the links 
of the nodes-links model for the diffusion cluster are of
this kind one has $\overline{d}=\ovd$. Indeed, we argued above
that the diffusion cluster {\it has} no dead ends or loops. 

We finally briefly discuss the temperature scaling of the 
dimensionless frequency in the DCA. 
As mentioned above, for homogeneous random walks on the diffusion cluster
$s\sgt\propto(s/\sigma)^{d_0/2}$ where $d_0=\ovd$. When this is
substituted into Eq.\ (\ref{22}) we find that the function
$f(\beta)$ in Eq.\ (\ref{22a}) is given by
$f(\beta)\propto\beta^{2/d_0}$. For $d_0=1.35$ one gets
$f\propto\beta^{1.48}$. In our simulations we find a similar power
law for $f(\beta)$, albeit with exponent $1.37$ \cite{sch99}.
Roling finds an exponent equal to $1.3$ \cite{rol01}.

\section{open questions}

Many questions about ac universality in the extreme disorder
limit remain unanswered:

\begin{itemize} 

\item For the random barrier model: How does the universal ac
conductivity depend on dimensionality? In particular: Is Eq.\
(\ref{21}) exact in 6 dimensions and above?

\item For the more general asymmetric hopping model, i.e., with
lattice sites of differing energies: Does this model also have ac
universality in the extreme disorder limit? If yes: Does the
universal ac conductivity depend on the choice of transition
rates? How does it depend on dimensionality?

\item More realistically one should consider hopping with only room for one
particle at each site. For ``Fermi hopping:'' Is there ac universality in
the extreme disorder limit? If yes: Is this the same as the ac universality
for the
random barrier model, which is the mean-field (Hartree) limit of
the Fermi model \cite{shk71,but74}?

\end{itemize}

In our opinion much remains to be done in this challenging field
of research. Compared to the 1970's and 1980's one now has the
possibility of extensive computer simulations at hand. This is
likely to bring further progress in the field.

\end{document}